\documentstyle[sprocl]{article}

\newcommand{\CP}{$CP$}

\begin{document}

\title{
\begin{flushright}
{\normalsize IIT-HEP-98/4\\
September 1998}
\vspace{0.1in}
\end{flushright}
SEARCHING FOR \CP\ VIOLATION IN PION DECAY\footnote
{Presented at the {\em Workshop on CP Violation},
3--8 July 1998, Adelaide, Australia.}}

\author{DANIEL M. KAPLAN}
\address{Physics, Illinois Institute of Technology,\\
Chicago, IL 60616, USA\\
E-mail: kaplan@fnal.gov}


\maketitle

\begin{abstract}
Surprisingly, until recently \CP\ violation in pion decay was not ruled out 
experimentally even at the percent level. I have derived the first 
experimental limit, $-0.01<A_{CP}<0.02$,
from old data on the anomalous magnetic moment $g-2$ of the 
muon. New data from the Brookhaven $g-2$ experiment might extend the search by 
a few orders of magnitude, but probably will not probe the 
theoretically-interesting sub-$10^{-7}$ regime.
\end{abstract}

\section{Introduction}

While the main topic of this Workshop is \CP\ violation in 
heavy-quark systems ($s$, $c$, $b$, and even $t$ have been discussed), it 
is perhaps surprising to realize that we don't even know whether {\em 
pion} decay conserves \CP! If we consider this question, we are likely to 
assume that the answer is yes, but an experimental test would be 
desirable.

\section{A Possible \CP-Violation Signature}

A conceptually-straightforward test is to 
measure the angular distribution of the electrons from the $\pi\to\mu\to 
e$ decay chain. This distribution is nonuniform due to parity violation in 
the weak interaction. \CP\ symmetry implies that the nonuniformity will be the
same
for electrons from $\pi^-$ as for positrons from $\pi^+$.\cite{Fieldetal}
This approach has the added benefit of being sensitive to \CP\ violation 
whether it occurs in pion decay or in muon decay.

The original observation\,\cite{Lederman-Garwin} of parity violation in this 
decay chain was carried out using positive pions stopping in carbon. A 
complementary experiment using $\pi^-$ would in principle allow a \CP\ 
test but is not feasible due to the bias introduced by negative-pion and 
muon capture in matter. It follows that pions decaying in flight in vacuum 
are required for such a \CP\ test.

\section{Muon $g-2$ Experiments}

An opportunity for such a measurement is afforded\,\cite{Kaplan,Morse} by 
the muon $g-2$ experiment. The anomalous 
magnetic moment $g-2$ of the muon has been measured with extreme 
precision using a muon storage ring at the CERN PS,\cite{Final} and an 
improved version of that apparatus is now in operation at the Brookhaven 
AGS.\cite{AGS} 

In a typical experiment, charged pions are injected 
into the ring, and muons from pion decay are captured in stable orbits. 
Electrons from the decay of the stored muons are detected, and, due to the 
muon polarization arising from parity violation, their rate in a given 
direction oscillates in time (see Fig.~1) as the muon spin precesses about the 
magnetic-field direction in the storage ring. The amplitude $A$ of this 
oscillation is proportional to both the degree of polarization of the muons 
and the analyzing power of the $\mu\to e$ decay. In principle both of 
these could be affected by phases arising from physics beyond the 
Standard Model, resulting in a detectable difference in oscillation 
amplitude between electrons from $\pi^-\to\mu^-$ and positrons from 
$\pi^+\to\mu^+$:
\begin{equation}
A_{CP} \equiv \frac{A_{\pi^+}-A_{\pi^-}}{A_{\pi^+}+A_{\pi^-}}\,.
\end{equation}

\begin{figure}
\centerline{\hspace{0.05in}\epsfxsize = 4in \epsffile{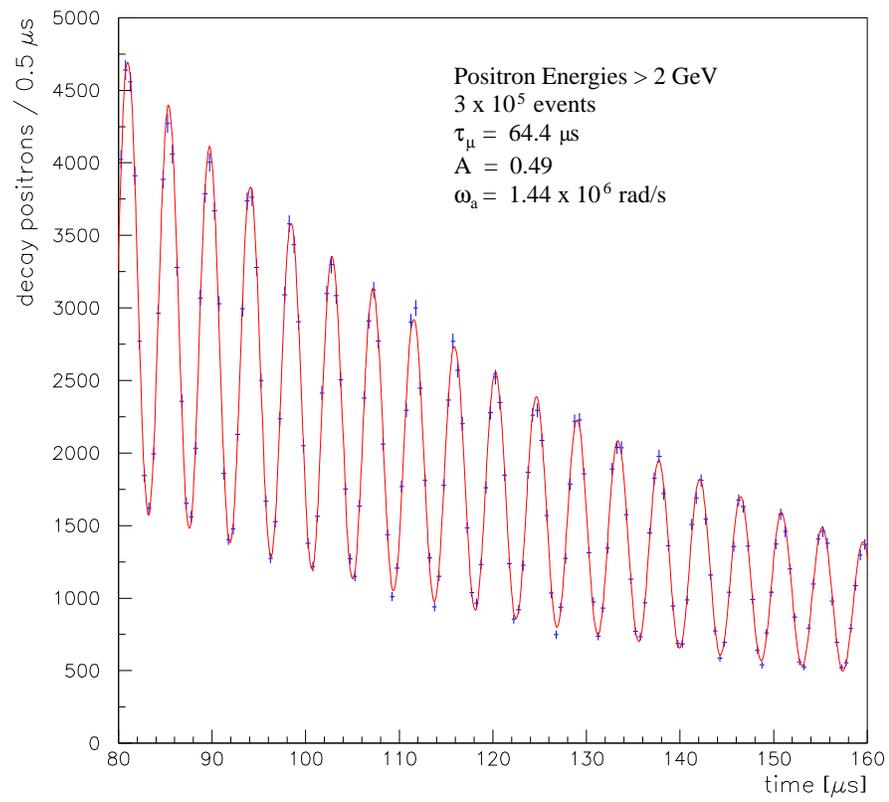}}
\caption{Example of data from the Brookhaven $g-2$ experiment showing
oscillations as described in text.}
\end{figure}

\section{The First Experimental Limit}

While the CERN $g-2$ group never pubished such a \CP\ limit, one has 
recently been derived from a figure in their ``Final Report on the CERN 
Muon Storage Ring,"\cite{Final} in which measurements are given of 
$A_{\pi^+}$ and $A_{\pi^-}$ for
four electron energy thresholds. The resulting limit is
\begin{equation}
-0.01<A_{CP}<0.02
\end{equation}
at 90\% confidence.\cite{Kaplan} Given the higher statistics, one might expect
${\cal O}(10^{-5})$ sensitivity to be possible in the Brookhaven experiment.
However, the CERN limit is dominated by systematic effects.\cite{Kaplan} It
remains to be seen whether these would dominate also at Brookhaven, and at what
level they can be controlled.

\section{Theoretical Upper Limit}

It is also of interest to estimate how large a \CP\ asymmetry is possible
theoretically. Discussions at this Workshop suggest that the effect is limited
to ${\cal O}(10^{-7})$ or less,\cite{He} since the final-state phase difference
necessary for an observable interference effect must here come from a weak
interaction involving a neutrino. Thus one has little expectation of an
observable effect at present levels of statistics. 

\section{Conclusions}

Given the
availability of pions in enormous numbers, the possibility of extremely precise
experiments in the future is not ruled out. It is also worth searching for 
alternative signatures to which new-physics contributions might be larger, for 
example $T$-odd correlations or measurements involving suppressed decay modes 
of the pion.

\section*{Acknowledgements}

I am grateful to D. Kawall and the Brookhaven $g-2$ collaboration for providing 
Fig.~1, and to S. Pakvasa for allowing me the opportunity to
squeeze this subject into the Workshop session that he chaired.
This work was supported in part by the U.S. Dept.\ of Energy and the  
Special Research Centre for the Subatomic Structure of Matter at the University 
of Adelaide.


\section*{References}
\begin{thebibliography}{9}

\bibitem{Fieldetal}
J. H. Field, E. Picasso, and F. Combley, {\it Sov.\ Phys.\ Usp.\ }{\bf 22}(4),
199 (1979).

\bibitem{Lederman-Garwin}
R. Garwin, L. M. Lederman, M. Weinrich,
{\it Phys.\ Rev.\ }{\bf 105}, 1415, (1957).

\bibitem{Kaplan}
D. M. Kaplan, {\it Phys.\ Rev.\ }D {\bf 57}, 3827 (1998).

\bibitem{Morse}
W. Morse, private communication; L. M. Lederman, private communication.

\bibitem{Final}
J. Bailey {\it et al.}, {\it Nucl.\ Phys.\ }B {\bf 150}, 1 (1979).

\bibitem{AGS}
B. L. Roberts {\it et al.}, ``Status of the New Muon ($g-2$) Experiment,"
in {\it Proc.\ 28th International Conference on High-Energy Physics 
(ICHEP 96)}, eds.\ Z. Ajduk and A. K. Wroblewski, World Scientific, 1997,
p.~1035.

\bibitem{He}
X.-G. He, S. Pakvasa, G. Valencia, {\it et al.}, private
communication.

\end{thebibliography}
\end{document}